# Jitter correction for asynchronous optical sampling terahertz spectroscopy using free-running pulsed lasers

**Mayuri Nakagawa\*, Natsuki Kanda, Toshio Otsu, Isao Ito, Yohei Kobayashi, and Ryusuke Matsunaga**

The Institute for Solid State Physics, The University of Tokyo, 5-1-5 Kashiwanoha, Kashiwa, Chiba 277-8581, Japan

\*Corresponding author: mayuri_nakagawa@issp.u-tokyo.ac.jp

**Abstract:** We demonstrate a jitter correction method for asynchronous optical sampling (ASOPS) terahertz (THz) time-domain spectroscopy using two free-running oscillators. This method simultaneously records the THz waveform and a harmonic of the laser repetition rate difference, $\Delta f\_r$, to monitor the jitter information for software jitter correction. By suppressing the residual jitter below 0.1 ps, the accumulation of the THz waveform is achieved without losing the measurement bandwidth. Our measurement of water vapor successfully resolves the absorption linewidths below 1 GHz, demonstrating a robust ASOPS with a flexible, simple, and compact setup without any feedback control or additional continuous-wave THz source.

1. Introduction

Asynchronous optical sampling (ASOPS) for terahertz time-domain spectroscopy (THz-TDS), which uses two pulsed lasers with slightly different repetition rates to generate and detect THz pulses, has realized high-resolution broadband spectroscopy and high scan rates without a mechanical delay stage [1]. The use of ASOPS has expanded the target of THz-TDS to substances with narrow spectral lines, such as gas molecules [2–4]. Because the diversity in the THz absorption spectra of volatile organic compound gases can be used as spectral fingerprints to distinguish each molecular species, high-resolution ASOPS THz-TDS is expected to be applicable for environmental safety [5]. In addition to gas molecules, ASOPS THz-TDS is also anticipated for broadband and time-resolved spectroscopy of sharp absorption lines in medicines [6] or biomolecules, such as DNA [7–10], to obtain broadband spectra and observe their time variation using a high scan rate.

Despite its potential advantages, ASOPS THz-TDS has not been fully utilized in society because of its inflexibility and technical difficulties. The most troublesome aspect is the timing jitter of the two pulsed lasers. Unless it is suppressed to less than the period of the THz wave, that is, approximately 1 ps, the THz-waveform data are distorted or canceled out by the accumulation processes. In ASOPS, the time delay in THz-TDS is expanded by a factor of $f_{r2}/\Delta f_r$ to the laboratory time, where $f_{r1}$ and $f_{r2}$ are the repetition rates of the gating and THz pulses, respectively, and $\Delta f_r \equiv f_{r2} - f_{r1}$ is the difference in the repetition rates. Therefore, the jitter is expressed by fluctuations in $f_{r2}$ and $\Delta f_r$. Considering the ratio between the fluctuations and mean values, the fluctuations in $\Delta f_r$ becomes more dominant in the jitter of the ASOPS. To overcome the jitter problem, typical ASOPS setups directly or indirectly stabilize $\Delta f_r$ by applying feedback control to the laser cavity length [2–4,11,12], the current of the laser diodes, or an electro-optic modulator [13–16]. In addition, $\Delta f_r$ can be stabilized by contriving cavity characteristics, such as using polarization-multiplexed, dual-wavelength, bidirectional modes of a single cavity [17–23], or using independent cavities with thermal and mechanical coupling [24,25]. Another option is to record the ASOPS signal in the sampling timings adjusted to the jitter, termed an "adaptive sampling" method [26]. This method enables more precise measurement than independently locking each repetition rate. However, an additional continuous-wave (CW) THz laser is necessary to create an adaptive clock. Although these ASOPS systems have successfully demonstrated high resolution and scan rates for THz spectroscopy, stabilization and special sampling schemes require expensive and/or sophisticated equipment, such as pulsed laser cavities capable of being stabilized with piezoelectric elements, special dual-laser configurations, or additional CW THz lasers. For practical applications, there is a high demand to devise a novel jitter-free ASOPS system with more flexibility, smaller equipment size, and lower cost than conventional systems.

In this study, we propose a jitter correction method for ASOPS THz-TDS using a software postprocessing algorithm. We provide a signal-generating apparatus and software procedure to calibrate the jitter and demonstrate jitter correction for measurements with free-running pulsed lasers. Despite a large fluctuation in the scan rate from 12–28 Hz in the 16-min measurement, the jitter is successfully corrected to allow accumulation of the ASOPS signal while maintaining a bandwidth in the THz frequency. We demonstrate the effectiveness of our jitter correction scheme by measuring low-pressure water vapor with high precision and spectral resolution, where linewidths of less than 1 GHz are resolved. Our method does not require any frequency stabilization, including mode-multiplexed lasers or feedback loops, and calibration is achieved using a simple circuit with a single radiofrequency (RF) signal generator without a CW laser. The simplified setup reduces cost and improves the robustness of the electrical system, which relies on high technical skills and a stable environment to achieve accurate long-term feedback control. Furthermore, the flexibility in choosing light sources will expand the purposes and targets of applying the ASOPS THz-TDS.

## 2. Sampling and analysis methods

The basic concept of the proposed jitter correction method is to acquire an RF signal with a harmonic frequency of $\Delta f_\mathrm{r}$, as well as a THz wave signal by ASOPS simultaneously, and correct the jitter on the software after recording the entire dataset. We employed the fourteenth harmonic ($14\Delta f_\mathrm{r}$) as the RF signal. In this study, we refer to this $14\Delta f_\mathrm{r}$ signal as the calibration signal and the THz waveform as the ASOPS signal. In an ideal case without any jitter or drift, the calibration signal is perfectly sinusoidal with a constant frequency of $14\Delta f_\mathrm{r}$. However, in the presence of jitter, the calibration signal is distorted, such that its phase is not proportional to the laboratory time. Our proposal is to calibrate the time axis by analyzing the phase of the calibration signal. The jitter in the ASOPS signal can be corrected by translating this phase into the time axis of the THz waveform. Note that the frequency-dependent phase shift in the electrical circuits also distorts the phase of the calibration signal. We consider the chirp effect and compensate for it in the software.

In this chapter, the optical and electronic systems used to acquire the ASOPS and calibration signals are explained in Section 2.1. The phase reconstruction scheme using the zero-crossing times of the calibration signal is discussed in Section 2.2. A software algorithm to correct the phase shift due to the chirp in the circuits and the jitter caused by the fluctuation of $\Delta f_\mathrm{r}$ is described in Sections 2.2 and 2.3. Finally, we describe the accumulation system to improve the signal-to-noise ratio (SNR) in Section 2.3.

*2.1. Data acquisition*

The pulsed-light sources in our setup are two Yb-doped fiber lasers with center wavelengths of approximately 1 μm, repetition rates of $f_\mathrm{r1} \sim f_\mathrm{r2} \sim 101$ MHz, and pulse durations of 70 and 90 fs. THz pulses are generated in a bias-free large-area emitter (TeraBlast, Protemics GmbH) and detected by electro-optic sampling in a (110) GaP crystal with balanced detection. This measurement yields the following three waveforms:

CH0. Interferogram (IFG) of the pump and gating lasers as a trigger signal.
CH1. RF signal with a varying frequency $14\Delta f_\mathrm{r}$ as the calibration signal.
CH2. Balanced detection signal of THz wave as the ASOPS signal.

The optical and electronic setups used in this method are shown in Fig. 1. The above three signals are acquired in CH0–CH2 with a sampling clock frequency of $2f_\mathrm{r1}$.

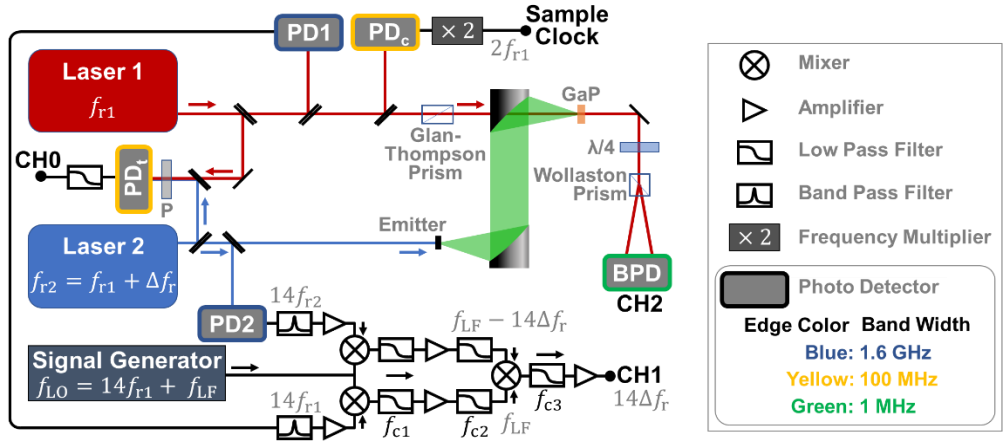

Fig. 1. Data acquisition setup. PD$_\mathrm{t}$: photodetector, BPD: balanced photodetector. P: polarizer, λ/4: quarter-wave plate. The edge colors of the photodetectors indicate their bandwidth.

The IFG detected by the photodetector (PD$_\mathrm{t}$) is filtered to cut the frequency components higher than $f_\mathrm{r}$ such that clear center bursts are extracted for use as triggers. The sampling clock signal is made from $f_\mathrm{r1}$ detected by PD$_\mathrm{c}$. The sampling clock repetition rate is electronically multiplied by a factor of two to be placed inside the clock range of the digitizer and divided by an integer in the software. The circuit shown at the bottom of Fig. 1 is used to generate the calibration signal $14\Delta f_r$. The detected signals of PD1 and PD2, including $nf_r$ ($n = 1, 2, ...$), are first filtered to pick up the frequency around $n = 14$. To extract the frequency related to $n = 14$ more clearly, they are converted to sufficiently low frequencies ($f_{LF}$ and $f_{LF} - 14\Delta f_r$) using a beat signal with the frequency of an RF signal generator $f_{LO}$. The beat between $f_{LF}$ and $f_{LF} - 14\Delta f_r$ is used as the calibration signal. The cutoff frequencies of the three-step low-pass filters, $f_{c1}, f_{c2}$, and $f_{c3}$ from left to right in the circuit, are determined to be $28f_r > f_{c1} = f_{c2} > f_{LF} > f_{c1}/2 > 7\Delta f_r$.

*2.2 Phase reconstruction of calibration signal*

To create a new time axis, the phase of the calibration signal is reconstructed from zero-crossing times. This process is illustrated in Fig. 2.

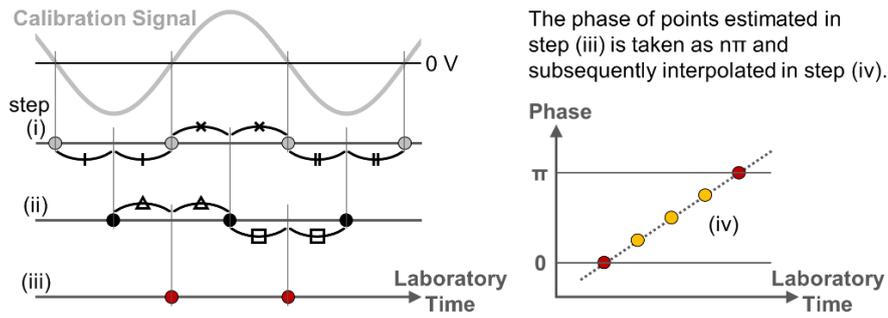

Fig. 2. Jitter correction algorism. Notations from (i) to (iv) indicate the steps of the phase reconstruction processes.

In the actual data, the calibration signal is slightly distorted by the nonlinearity of the last amplifier circuit. Therefore, the process must be carefully developed such that it is not affected by the nonuniformity between the falling and rising zero-crossing times. First, we detect the zero-crossing points of rising and falling at laboratory time (i). The middle points of the zero-crossing points are subsequently determined, as shown in (ii). The spacings of points (ii) are impartial; however, the origin offset still contains a bias. By further setting the middle point of (ii), as shown in (iii), the offset bias can be eliminated. The phases in (iii) are defined as $0, \pi, 2\pi$ .... The intermediate phases between points (iii) are subsequently reconstructed using linear interpolation (iv).

*2.3 Chirp correction in circuits*

Most parts of the jitter can be corrected using the scheme in Sections 2.1 and 2.2. The fluctuation of frequency $14\Delta f_r$ also induces jitter originating from a frequency-dependent phase shift or "chirp" in electronic circuits, such as the final low-pass and amplifier. Here, this chirped delay is compensated for by comparing the recorded calibration signal with the IFG. We use the burst timings of the IFG at every $1/\Delta f_r$ as the trigger timings of the signal accumulation. In a chirp-free situation, the phases determined in Section 2.2 should be the same at every trigger timing, except for constant increments ($14 \times 2\pi$). Therefore, we can inspect the phases at the trigger timings to check the chirp effect. The frequencies at each point can also be determined using phase spacing in the time domain. Using a phase-frequency curve determined by fitting these data, the phase delay of the entire dataset can be determined from the frequency variation over the laboratory time. This phase delay is ultimately subtracted from the phase determined in Section 2.2.

*2.4 Jitter correction and accumulation*

The entire jitter, including the chirp effect, are corrected simultaneously by translating the phase determined in Section 2.3 to a new time axis for the THz waveform. In particular, the trigger intervals are $14 \times 2\pi$ along the phase axis and $1/f_{r2}$ along the new time axis. Therefore, the phase on each acquired data point is rescaled as a time axis by the division of $14 \times 2\pi \times f_{r2}$.

To improve the SNR, we accumulated the signal after jitter correction. To accumulate data on the same timings, the corrected data are interpolated to provide a time axis with equal intervals. The interpolated data are cut on trigger timings and averaged. Note that if the jitter is not corrected, the signals would cancel each other out during accumulation, particularly at high frequencies. To maintain the signal strength over a broad bandwidth, the accumulation must be processed after jitter correction.

## 3. Experiments and discussion

*3.1 Examination on jitter correction and signal accumulation*

We measure the ASOPS signal to examine the effect of the jitter correction. In Fig. 3(a), the raw data for CH0 to CH2 at the start of the measurement are shown with an offset. Owing to the inevitable drift and fluctuation of the free-running oscillators, the $\Delta f_r$ evaluated from the recorded calibration signal varies from 12 to 28 Hz during the 16-min measurement. To clarify the effect of jitter correction, the ASOPS signals before jitter correction are shown in Fig. 3(b). The dark and light-green data indicate the first and last scans used for accumulation, respectively. Although both signals have the same trigger, the burst timings of the THz pulses are different from each other owing to the drift of $\Delta f_r$. The jitter and drift in the recorded data are corrected using the method described in Section 2. To set the pulse-to-pulse time range of the corrected time axis to $1/f_{r2}$, an approximate frequency $f_{r2} = 100.78$ MHz is used for analysis.

To determine the chirp effect in electronics, the relationship between the phase delay and frequency of the calibration signal is obtained from this measurement, as shown in Fig. 3(d). The black circles indicate the data for each trigger timing. This phase delay is estimated by taking the remainder obtained by dividing the phase at the trigger by $14 \times 2\pi$. The reduced points emphasized by the white circles in Fig. 3(d) are used for the fitting. The fitted quadratic curve, shown as a cyan line, is used to determine the phase delay at each data point. The ASOPS signals after jitter correction are shown in Fig. 3(c). The burst timings of the THz signals are successfully matched after the correction.

The residual jitter is quantitatively estimated as follows: Because it is difficult to directly estimate the jitter in the ASOPS THz signal owing to noise, the jitter between the calibration signal and trigger timings after the correction is determined. Specifically, the phase of the calibration signal after chirp subtraction on trigger timings is divided by $14 \times 2\pi$ and the remainder is divided by $14 \times 2\pi \times f_{r2}$ to translate it to time. The residual jitter is the time variation between the trigger and the next trigger, which is plotted in Fig. 3(e). The root-mean square of the entire measurement is 0.09 ps. This guarantees that the residual jitter after correction does not harm the bandwidth for accumulating the THz wave signal.

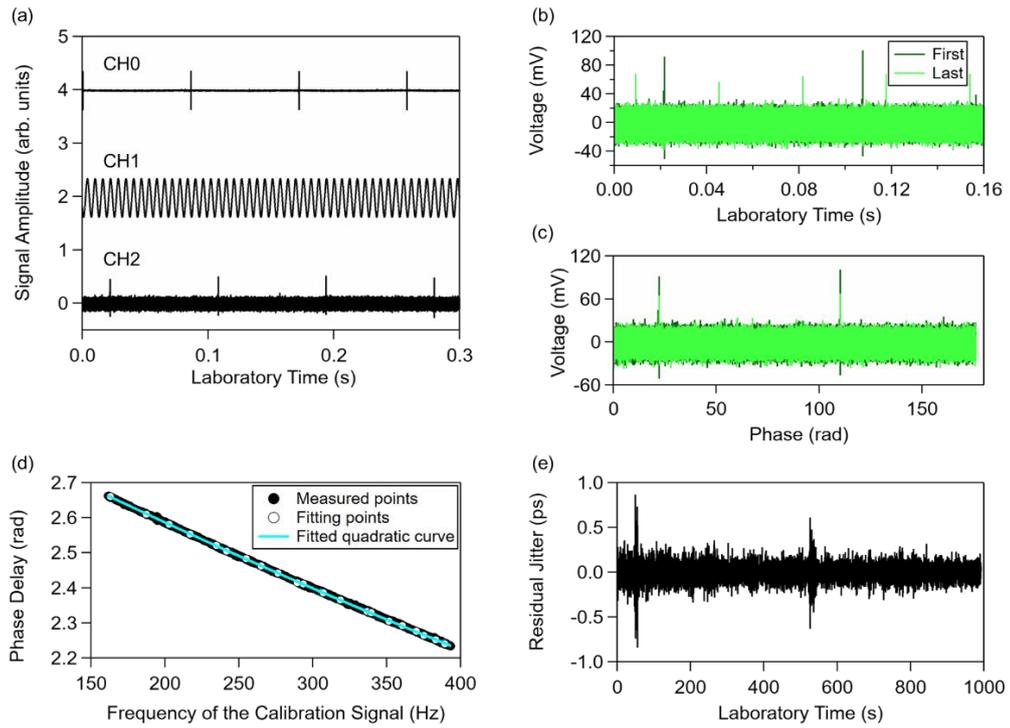

Fig. 3. Process and the effect of the jitter correction. (a) Acquired raw data. (b) Raw ASOPS signal at the first and the last scans of the data acquisition. (c) Corrected ASOPS signals at the first and the last scans. (d) Fitting to determine chirp of electronics. (e) Residual jitter at each scan.

The signal accumulation is executed by two pulse-to-pulse intervals 10,970 times within a 16-min measurement. The time-domain waveforms before and after averaging are shown in Fig. 4(a). The green data indicate signals before accumulation. The averaged waveforms are plotted as blue curves. For each data point, waveforms with short (40 ps) and long (20 ns) time ranges are shown with light and vivid colors, respectively. After the accumulation, the noise is reduced, but the sharp bursts of the THz pulse hold their waveform. The power spectra obtained from Fig. 4(a) are presented in Fig. 4(b).

In Fig. 4(b), the signal is hardly visible in the spectrum for a longer time window before accumulation (light green). After the accumulation, the signal can be observed because the noise floor in the high-frequency range is suppressed by a factor of approximately 10,000, corresponding to an averaging number of 10,970 (light blue). A shorter time window is also effective for obtaining a smaller noise floor, which is a trade-off with a worse spectral resolution. The noise floor is suppressed approximately

5000 times, which corresponds to the ratio of the lengths of the time window. Without accumulation, a THz signal of up to 1 THz is observed above the noise floor (vivid green). The noise floor is further suppressed by the accumulation, which enables detection up to 2 THz (vivid blue). The vivid green and blue spectra in Fig. 3(b) show good agreement at frequencies below 1 THz. These results confirm the successful jitter-corrected accumulation with a large bandwidth in the THz frequency range.

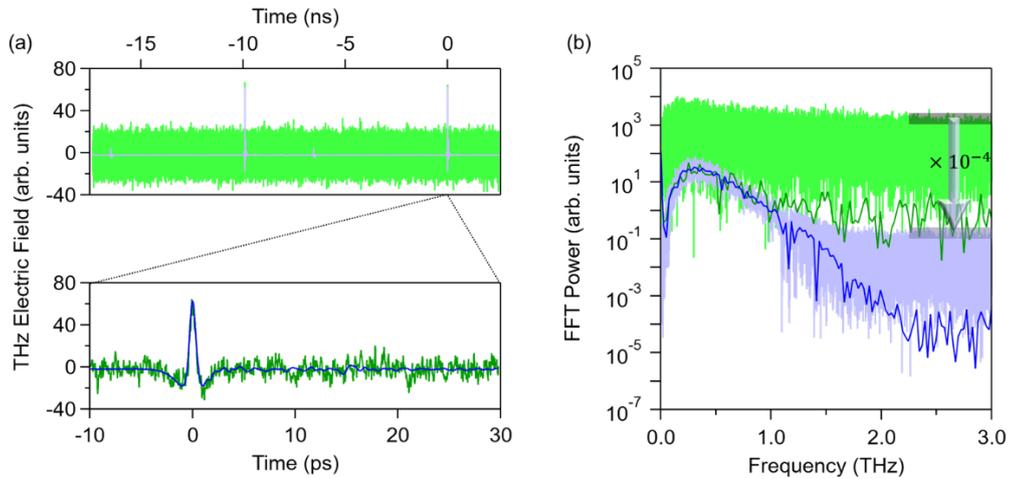

Fig. 4. ASOPS signal with the jitter correction. (a)(b) Time-domain waveforms and power spectra, for one scan (green) and 10,970 scans (blue), respectively. Light (vivid) color corresponds to data with the time window of 40 ps (20 ns).

*3.2 Measurement of water vapor*

To demonstrate the fine frequency resolution, we examine the measurement of water vapor. For a quantitative evaluation of the optical density, a gas cell is vacuumed and then filled with saturated water vapor using the setup shown in Fig. 5(a). In this situation, the absorption lines of water become sharper than those at atmospheric pressure [27]. The measurement is conducted at 24.8 °C, where the saturated water vapor pressure is estimated to be 3131 Pa. The entire THz beam path is purged with dry nitrogen gas. The length of the gas cell is 168.2 mm, which are enclosed on both sides with two polytetrafluoroethylene (PTFE) windows with a thickness of 3.2 mm.

During this measurement, $\Delta f_r$ is determined from the recorded calibration signal and varies from 19.5 to 24.7 Hz due to the drift. $f_{r1}$ is also recorded by a frequency counter with a gate time of 1 s. Then, $f_{r2}$ is retrieved from these $\Delta f_r$ and $f_{r1}$. The recorded $f_{r1}$ and retrieved $f_{r2}$ values are shown in Fig. 5(b). To improve the accuracy of the corrected time axis, the mean value of the varying $f_{r2}$, 100,782,386 Hz, is used to calibrate the pulse-to-pulse interval to $1/f_{r2}$. Because the variation in $f_{r2}$ is 8.5 Hz during this measurement, the variation in the frequencies of the measured THz combs at 1 THz is approximately $8.5 \text{ Hz} \times (1 \text{ THz}/f_{r2}) = 84 \text{ kHz}$ which is sufficiently smaller than the frequency resolution $f_{r2}$.

The measured power spectrum is shown as a black curve in the left panel of Fig. 5(c). The narrow dips caused by the absorption of water vapor are shown in the right panels. The absorption lines calculated from the HITRAN database [28] are also shown as blue curves in Fig. 5(c), with an offset. In this calculation, the linewidth and absorption strength are quantitatively evaluated by considering the pressure, temperature, and propagation length. The widths of the calculated absorption lines are 0.90, 0.85, and 0.91 GHz, respectively. The measured spectra and the database show good agreement in terms of both the absorption intensity and linewidth. These results confirm the high-frequency resolution of our jitter-corrected ASOPS.

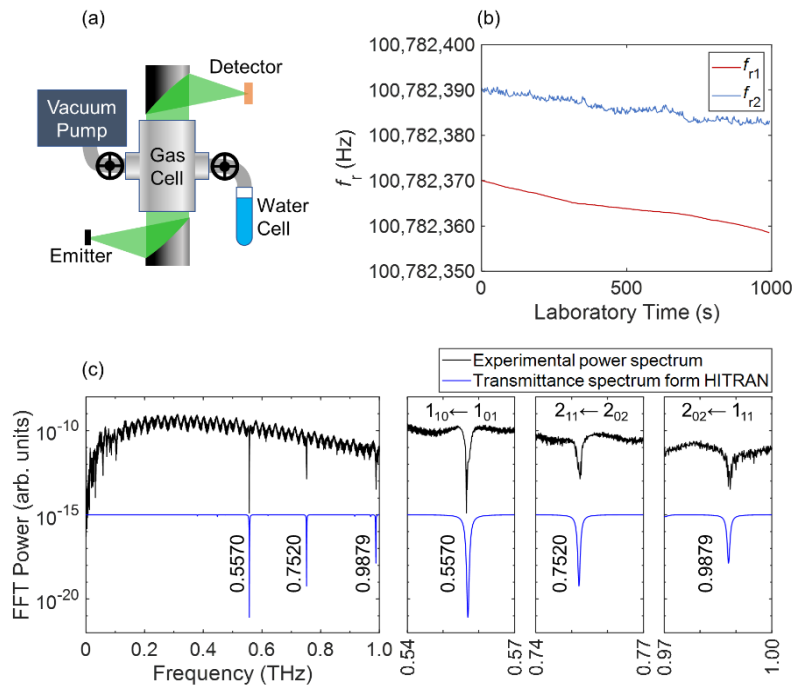

Fig. 5. (a) Schematic of setup with the water vaper cell. (b) Variation of the repetition frequency during the measurement on water vapor. (c) Spectra of the THz pulses with the absorption lines of water vapor. Assigned transitions are indicated on expended figures [29,30].

*3.3 Discussion*

In this section, we discuss the validity of this method and its differences compared to the conventional ASOPS. Our method assumes that jitter in the ASOPS is dominated by the fluctuation of $\Delta f_\mathrm{r}$. In this context, the concept of our method is similar to that of previous studies on frequency stabilization by locking $\Delta f_\mathrm{r}$ or its harmonics [2,11]. However, locking $\Delta f_\mathrm{r}$ is sometimes difficult to achieve in the presence of large fluctuations owing to the tolerance of the feedback mechanisms. Therefore, to the best of our knowledge, previous ASOPS systems with locking of $\Delta f_\mathrm{r}$ have locked $f_\mathrm{r}$ in addition to $\Delta f_\mathrm{r}$.

In these locking schemes, error signals for the feedback loop are created by low-pass filtering the frequency-mixed signal of the RF reference and $n\Delta f_\mathrm{r}$. The bandwidth of the frequency stabilization is limited by the stability of the RF signal generation and the cutoff of the low-pass filter. In contrast, our method directly monitors the harmonic of $\Delta f_\mathrm{r}$ instead of generating an error signal. In addition, the fluctuation of the RF oscillation $f_\mathrm{LO}$ is passively canceled out when taking the beat of $f_\mathrm{LF}$ and $f_\mathrm{LF} - 14\Delta f_\mathrm{r}$, although we use an RF signal generator. Therefore, our method is not limited by the stability of RF reference or low-pass filters. Clear detection of $14\Delta f_\mathrm{r}$ using this RF local oscillator close to $14f_\mathrm{r}$ suppresses fast jitter, in principle up to a bandwidth of $14\Delta f_\mathrm{r}/2$. Because the stabilization-free jitter correction has a greater tolerance to fluctuations of $\Delta f_\mathrm{r}$, our method can be applied to independent free-running oscillators.

Most ASOPS THz-TDS techniques use IFG, sum-frequency generation, or two-photon absorption of pulsed lasers as a trigger for signal accumulation [3,11,12,18,31]. However, using such triggers fixes the jitter only around the triggered time for every $1/\Delta f_\mathrm{r}$. The change in the rescaling factor $\Delta f_\mathrm{r}/f_\mathrm{r2}$ from the laboratory time is also ignored, which results in larger jitters as the data are away from the triggered time. By using the intervals and phases of the IFG, the jitters at the entire time window between the triggers can be corrected using software algorithms, as previously studied in the scheme of dual-comb spectroscopy [33–36]. However, in these methods, the tracking speed of the jitter is limited to the repetition frequency of IFG $\Delta f_\mathrm{r}$ [37]. To correct faster jitter, an enhancement in the tracking speed is essential. Recording $n\Delta f_\mathrm{r}$ signal can improve the speed, as reported in dual-comb spectroscopy [38,39].

Our postprocessing algorithms in ASOPS could enhance the tracking speed by up to $14\Delta f_r$, which overcomes the limitation $\Delta f_r$ in the methods of jitter correction with IFG. This is because we separately detect the respective lasers with comb spacing $f_r$, in contrast to the detection of IFG, where the comb spacing of $\Delta f_r$ limits the jitter tracking rate. We increased the frequency of this limitation to $f_r$ which is much higher than $\Delta f_r$. Furthermore, information about the carrier-envelope offset frequency is unnecessary for the ASOPS THz-TDS. Directly detecting the respective lasers by PDs is also advantageous in removing the effect of carrier-envelope offset frequency because of the slow response speed of PDs. Therefore, our system simplifies the observation of jitter in ASOPS THz-TDS.

As another way to achieve a sufficient bandwidth for the ASOPS without frequency stabilization, adaptive sampling is known to suppress jitter by using $n$-th harmonic of $\Delta f_r$ as a sampling clock, where $n \sim f_{CW}/f_r$. To apply this method to ASOPS THz-TDS, the beat signal between $n$-th mode of the photoconductive (PC)-THz comb and THz CW laser is first obtained, and then the beat signal for each pulsed laser is used to generate a high-frequency adaptive clock. It has a much faster jitter tracking rate than our method and can correct the jitter in real time; however, it requires a THz CW laser with a frequency $f_{CW}$ and additional photoconductive antennas to create high-frequency harmonics. In addition, a fast jitter corresponding to phase noise with a high offset frequency usually does not have a large influence on the amount of jitter. The advantage of our method is that it adequately corrects the jitter with a minimal necessary high-frequency reference, which can be easily created. Our system consists of independent free-running lasers, and a single RF signal generator without a CW laser is much easier to provide than any other low-jitter setup.

To achieve further jitter suppression, $n\Delta f_r$ to calibrate jitter should be a higher frequency. It is possible by changing the numbering of the calibration signal $n = 14$ by altering $f_{LO}$ and filter frequencies, or by increasing $\Delta f_r$. In the present system, undesirable harmonics due to amplification and their different phase delays hamper accurate phase detection at nonzero voltages. In the future, amplification with less distortion would suppress the anharmonicity of the calibration signal. Such a less distorted signal would allow us to use the entire phase of the calibration signal, which could correct even a much faster jitter.

To improve the SNR, lasers with higher $f_r$ would be a better solution to shorten the excessive time window, which becomes noisier farther from the burst of the THz pulse. Although a shorter time window results in worse frequency resolution, the resolution is sufficiently fine for most practical applications. A faster scan rate with a shorter time window facilitates frequent accumulation. A higher $f_r$ also helps to lower the measurement speed of the ASOPS signal and, consequently, secure the measurement bandwidth.

An expected advantage of our method for application is that we can choose pulsed light sources from a wide variety of options. Our method does not require controlling mechanism in laser resonator by a piezo element, which allows freely choosing lasers in terms of center frequency, pulse width, and repetition frequency, and even the commercially available free-running Ti:Sapphire lasers can be used for ASOPS experiments. Because this method is also expected to be robust against environmental changes, a long-term measurement, even outdoors, might be possible using our method. Using the smallness of our system and the rapidity of the ASOPS, a handy THz-TDS with high-frequency resolution may also be achieved. Our concept of monitoring $n\Delta f_r$ can also be applied to other new techniques. For example, the spectrally interleaving ASOPS, which improves the frequency resolution by sweeping $f_r$ [40], can be enhanced by our method. The resolution of this method is determined by the step width of the sweeping $f_r$. If our method is applied, $f_r$ can be swept continuously by suppressing the jitter. Our method is also expected to solve the timing jitter problem in electronically controlled optical sampling which modulates $f_r$ of one of the two pulsed lases [41].

## 4. Conclusions

We demonstrated a new jitter correction method that corrects the ASOPS signal on software using the fourteenth harmonic of the repetition rate difference. The residual jitter after correction was suppressed to less than 0.1 ps even though free-running oscillators are used. This adequate jitter suppression allowed us to accumulate an ASOPS signal while maintaining the measurement bandwidth and resolution. The evaluation of residual jitter using IFG and the successful signal averaging after correcting the jitter demonstrate the potentially broadband THz-TDS of this method. The frequency resolution of the ASOPS was confirmed by measuring a low-pressure gas cell with sharp absorption peaks that were narrower than 1 GHz. Our method increases the flexibility of the ASOPS system from the perspective of size, cost, robustness, and choices of light sources. In particular, only one RF signal generator is required for this measurement in contrast to conventional schemes. In addition, CW laser sources are not required for the reference frequency. Importantly, this method can be applied to independent pulsed lasers with abrupt temperature changes. These merits increase the potential of the ASOPS to choose the most suitable frequency and pulse characteristics for the measurement targets, as well as a portable and robust measurement system. Our method can be applied in many fields, such as medicine, structural inspection, and various pump-probe ASOPS for frequencies other than THz.

**Funding:** JSPS KAKENHI (JP22K18269) and JST PREST (JPMJPR20LA and JPMJPR2006).

**Disclosures:** The authors declare no conflicts of interest.

**Data availability:** Data for this study are available from the corresponding authors upon reasonable request.

**References:**

1. T. Yasui, E. Saneyoshi, and T. Araki, "Asynchronous optical sampling terahertz time-domain spectroscopy for ultrahigh spectral resolution and rapid data acquisition," Appl. Phys. Lett. **87**, 061101 (2005). https://doi.org/10.1063/1.2008379.

2. G. Klatt, R. Gebs, H. Schäfer, M. Nagel, C. Janke, A. Bartels, and T. Dekorsy, "High-Resolution Terahertz Spectrometer," IEEE J. Sel. Top. Quantum Electron. **17**(1), 159–168 (2011). https://doi.org/10.1109/jstqe.2010.2047635.

3. T. Yasui, K. Kawamoto, YD. Hsieh, Y. Sakaguchi, M. Jewariya, H. Inaba, K. Minoshima, F. Hindle, and T. Araki, "Enhancement of spectral resolution and accuracy in asynchronous-optical-sampling terahertz time-domain spectroscopy for low-pressure gas-phase analysis," Opt. Express **20**(14), 15071–15078 (2012). https://doi.org/10.1364/oe.20.015071.

4. Y.-D. Hsieh, S. Nakamura, D. G. Abdelsalam, T. Minamikawa, Y. Mizutani, H. Yamamoto, T. Iwata, F. Hindle and T. Yasui, "Dynamic terahertz spectroscopy of gas molecules mixed with unwanted aerosol under atmospheric pressure using fibre-based asynchronous-optical-sampling terahertz time-domain spectroscopy," Sci. Rep. **6**, 28114 (2016). https://doi.org/10.1038/srep28114.

5. V. Galstyan, A. D'Arco, M. D. Fabrizio, N. Poli, S. Lupi, and E. Comini, "Detection of volatile organic compounds: From chemical gas sensors to terahertz spectroscopy," Rev. Anal. Chem. **40**(1), 33–57 (2021). https://doi.org/10.1515/revac-2021-0127.

6. T. Sasaki, T. Sakamoto, and M. Otsuka, "Sharp Absorption Peaks in THz Spectra Valuable for Crystal Quality Evaluation of Middle Molecular Weight Pharmaceuticals," J. Infrared Millim. Terahertz Waves **39**, 828–839 (2018). https://doi.org/10.1007/s10762-018-0494-2.

7. B. M. Fischer, M. Walther, and P. U. Jepsen, "Far-infrared vibrational modes of DNA components studied by terahertz time-domain spectroscopy," Phys. Med. Biol. **47**, 3807 (2002). https://doi.org/10.1088/0031-9155/47/21/319.

8. T. Globus, D. Woolard, T. W. Crowe, T. Khromova, B. Gelmont, and J. Hesler, "Terahertz Fourier transform characterization of biological materials in a liquid phase," J. Phys. D: Appl. Phys. **39**, 3405 (2006). https://doi.org/10.1088/0022-3727/39/15/028.

9. W. Zhang, E. R. Brown, M. Rahman, and M. L. Norton, "Observation of terahertz absorption signatures in microliter DNA solutions," Appl. Phys. Lett. **102**, 023701 (2013). https://doi.org/10.1063/1.4775696.

10. X. Yang, X. Zhao, K. Yang, Y. Liu, Y. Liu, W. Fu, and Y. Luo, "Biomedical applications of terahertz spectroscopy and imaging," Trends Biotechnol. **34**(10), 810–824 (2016). https://doi.org/10.1016/j.tibtech.2016.04.008.

11. G. Klatt, R. Gebs, C. Janke, T. Dekorsy, and A. Bartels, "Rapid-scanning terahertz precision spectrometer with more than 6 THz spectral coverage," Opt. Express **17**, 22847–22854 (2009). https://doi.org/10.1364/oe.17.022847.

12. J. T. Good, D. B. Holland, I. A. Finneran, P. B. Carroll, M. J. Kelley, and G. A. Blake, "A decade-spanning high-resolution asynchronous optical sampling terahertz time-domain and frequency comb spectrometer," Rev. Sci. Instrum. **86**, 103107 (2015). https://doi.org/10.1063/1.4932567.

13. M. Oeri, O. Peters, M. Wolferstetter, and R. Holzwarth, "Compact, high-speed sampling engine for pulsed femtosecond lasers," Proc. SPIE **11348**, 1134808 (2020). https://doi.org/10.1117/12.2558423.

14. M. Nakagawa, M. Okano, and S. Watanabe, "Polarization-sensitive terahertz time-domain spectroscopy system without mechanical moving parts," Opt. Express **30**(16), 29421–29434 (2022). https://doi.org/10.1364/oe.460259.

15. M. Okano and S. Watanabe, "Triggerless data acquisition in asynchronous optical-sampling terahertz time-domain spectroscopy based on a dual-comb system," Opt. Express **30**(22), 39613–39623 (2022). https://doi.org/10.1364/oe.472192.

16. N. Surkamp, B. Döpke, C. Brenner, K. Orend, C. Baer, T. Musch, T. Prziwarka, A. Klehr, A. Knigge, M. R. Hofmann, "Mode-locked diode lasers for THz asynchronous optical sampling," Proc. SPIE **10917**, 109171C (2019). https://doi.org/10.1117/12.2508396.

17. X. Zhao, G. Hu, B. Zhao, C. Li, Y. Pan, Y. Liu, T. Yasui, and Z. Zheng,

"Picometer-resolution dual-comb spectroscopy with a free-running fiber laser," Opt. Express **24**, 21833–21845 (2016). https://doi.org/10.1364/oe.24.021833.

18. G. Hu, T. Mizuguchi, R. Oe, K. Nitta, X. Zhao, T. Minamikawa, T. Li, Z. Zheng, and T. Yasui. "Dual terahertz comb spectroscopy with a single free-running fibre laser," Sci. Rep. **8**, 11155 (2018). https://doi.org/10.1038/s41598-018-29403-9.
19. A. E. Akosman and M. Y. Sander, "Dual comb generation from a mode-locked fiber laser with orthogonally polarized interlaced pulses," Opt. Express **25**, 18592–18602 (2017). https://doi.org/10.1364/oe.25.018592.
20. B. Willenberg, J. Pupeikis, L. M. Krüger, F. Koch, C. R. Phillips, and U. Keller, "Femtosecond dual-comb Yb:CaF$_2$ laser from a single free-running polarization-multiplexed cavity for optical sampling applications," Opt. Express **28**, 30275–30288 (2020). https://doi.org/10.1364/oe.403072.
21. S. Mehravar, R. A. Norwood, N. Peyghambarian, and K. Kieu, "Real-time dual-comb spectroscopy with a free-running bidirectionally mode-locked fiber laser," Appl. Phys. Lett. **108**, 231104 (2016). https://doi.org/10.1063/1.4953400.
22. T. Ideguchi, T. Nakamura, Y. Kobayashi, and K. Goda, "Kerr-lens mode-locked bidirectional dual-comb ring laser for broadband dual-comb spectroscopy," Optica **3**, 748–753 (2016). https://doi.org/10.1364/optica.3.000748.
23. R. D. Baker, N. T. Yardimci, Y.-H. Ou, K. Kieu, and M. Jarrahi, "Self-triggered asynchronous optical sampling terahertz spectroscopy using a bidirectional mode-locked fiber laser," Sci. Rep. **8**, 14802 (2018). https://doi.org/10.1038/s41598-018-33152-0.
24. G. Villares, J. Wolf, D. Kazakov, M. J. Süess, A. Hugi, M. Beck, and J. Faist, "On-chip dual-comb based on quantum cascade laser frequency combs," Appl. Phys. Lett. **107**(25), 251104 (2015). https://doi.org/10.1063/1.4938213.
25. Y. Nakajima, Y. Kusumi, and K. Minoshima, "Mechanical sharing dual-comb fiber laser based on an all-polarization-maintaining cavity configuration," Opt. Lett. **46**, 5401–5404 (2021). https://doi.org/10.1364/ol.440818.
26. T. Yasui, R. Ichikawa, Y.-D. Hsieh, K. Hayashi, H. Cahyadi, F. Hindle, Y. Sakaguchi, T. Iwata, Y. Mizutani, H. Yamamoto, K. Minoshima and H. Inaba, "Adaptive sampling dual terahertz comb spectroscopy using dual free-running femtosecond lasers," Sci. Rep. **5**, 10786 (2015). https://doi.org/10.1038/srep10786.
27. P. F. Bernath, "Pressure broadening," in *Spectra of atoms and molecules* (Oxford University Press, 2005), pp. 28-29.
28. I. E. Gordon, L. S. Rothman, R. J. Hargreaves, R. Hashemi, E. V. Karlovets, F. M. Skinner, E. K. Conway, C. Hill, R. V. Kochanov, Y. Tan, P. Wcisło, A. A. Finenko, K. Nelson, P. F. Berna, M. Birk, V. Boudon, A. Campargue, K. V. Chance, A. Coustenis, B. J. Drouin, J.-M. Flaud, R. R. Gamache, J. T. Hodges, D. Jacquemart, E. J. Mlawer, A. V. Nikitin, V. I. Perevalov, M. Rotger, J. Tennyson, G. C. Toon, H. Tran, V. G. Tyuterev, E. M. Adkins, A. Baker, A. Barbe, E. Canè, A. G. Császár, A. Dudaryonok, O. Egorov, A. J. Fleisher, H. Fleurbaey, A. Foltynowicz, T. Furtenbacher, J. J. Harrison, J.-M. Hartmann, V.-M. Horneman,

X. Huang, T. Karman, J. Karns, S. Kassi, I. Kleiner, V. Kofman, F. Kwabia-Tchana, N. N. Lavrentieva, T. J. Lee, D. A. Long, A. A. Lukashevskaya, O. M. Lyulin, V. Yu. Makhnev, W. Matt, S. T. Massie, M. Melosso, S. N. Mikhailenko, D. Mondelain, H. S. P. Müller, O. V. Naumenko, A. Perrin, O. L. Polyansky, E. Raddaoui, P. L. Raston, Z. D. Reed, M. Rey, C. Richard, R. Tóbiás, I. Sadiek, D. W. Schwenke, E. Starikova, K. Sung, F. Tamassia, S. A. Tashkun, J. Vander Auwera, I. A. Vasilenko, A. A. Vigasin, G. L. Villanueva, B. Vispoel, G. Wagner, A. Yachmenev, and S. N. Yurchenko, "The HITRAN2020 molecular spectroscopic database," J. Quant. Spectrosc. Radiat. Transfer **277**, 107949 (2022). https://doi.org/10.1016/j.jqsrt.2021.107949.

29. T. Seta, H. Hoshina, Y. Kasai, I. Hosako, C. Otani, S. Loßow, J. Urban, M. Ekström, P. Eriksson, and D. Murtagh, "Pressure broadening coefficients of the water vapor lines at 556.936 and 752.033GHz," J. Quant. Spectrosc. Radiat. Transfer **109**(1), 144–150 (2008). https://doi.org/10.1016/j.jqsrt.2007.06.004.

30. H. Hoshina, T. Seta, T. Iwamoto, I. Hosako, C. Otani, and Y. Kasai, "Precise measurement of pressure broadening parameters for water vapor with a terahertz time-domain spectrometer," J. Quant. Spectrosc. Radiat. Transfer **109**(12–13), 2303-2314 (2008). https://doi.org/10.1016/j.jqsrt.2008.03.005.

31. H. Zhang, B. Su, X. Yang, Y. Wu, J. He, C. Zhang, and D. R. Jones, "Asynchronous optical sampling data-acquisition trigger-signal derived from pulse coherence coincidence," Rev. Sci. Instrum. **89**, 113108 (2018). https://doi.org/10.1063/1.5051072.

32. I. Coddington, W. C. Swann, and N. R. Newbury, "Coherent Multiheterodyne Spectroscopy Using Stabilized Optical Frequency Combs," Phys. Rev. Lett. **100**, 013902 (2008). https://doi.org/10.1103/physrevlett.101.049901.

33. D. Burghoff, Y. Yang, and Q. Hu, "Computational multiheterodyne spectroscopy," Sci. Adv. **2**(11), e1601227 (2016). https://doi.org/10.1126/sciadv.1601227.

34. N. B. Hébert, J. Genest, J.-D. Deschênes, H. Bergeron, G. Y. Chen, C. Khurmi, and D. G. Lancaster, "Self-corrected chip-based dual-comb spectrometer," Opt. Express **25**, 8168–8179 (2017). https://doi.org/10.1364/oe.25.008168.

35. H. Yu, K. Ni, Q. Zhou, X. Li, X. Wang, and G. Wu, "Digital error correction of dual-comb interferometer without external optical referencing information," Opt. Express 27, 29425–29438 (2019). https://doi.org/10.1364/oe.27.029425.

36. H. Yu, Y. Li, Q. Ma, Q. Zhou, X. Li, W. Ren, and K. Ni, "A coherent-averaged dual-comb spectrometer based on environment-shared fiber lasers and digital error correction," Opt. Laser Technol. 156, 108498 (2022). https://doi.org/10.1016/j.optlastec.2022.108498.

37. N. B. Hébert, V. Michaud-Belleau, J. -D. Deschênes and J. Genest, "Self-Correction Limits in Dual-Comb Interferometry," IEEE J. Quantum Electron. **55**(4), 8700311, 1–11 (2019). https://doi.org/10.1109/jqe.2019.2918935.

38. O. Kara, Z. Zhang, T. Gardiner, and D. T. Reid, "Dual-comb mid-infrared spectroscopy with free-running oscillators and absolute optical calibration from a radio-frequency reference," Opt. Express **25**, 16072-16082 (2017). https://doi.org/10.1364/oe.25.016072.

39. Z. Zhu, K. Ni, Q. Zhou, and G. Wu, "Digital correction method for realizing a phase-stable dual-comb interferometer," Opt. Express **26**, 16813-16823 (2018). https://doi.org/10.1364/oe.26.016813.
40. Y.-D. Hsieh, Y. Iyonaga, Y. Sakaguchi, S. Yokoyama, H. Inaba, K. Minoshima, F. Hindle, T. Araki, and T. Yasui, "Spectrally interleaved, comb-mode-resolved spectroscopy using swept dual terahertz combs," Sci. Rep. **4**, 3816 (2014). https://doi.org/10.1038/srep03816.
41. Y. Kim and D.-S. Yee, "High-speed terahertz time-domain spectroscopy based on electronically controlled optical sampling," Opt. Lett. **35**, 3715–3717 (2010). https://doi.org/10.1364/ol.35.003715.